\documentclass[12pt]{article}
\usepackage{amssymb}
\usepackage{amsmath}
\usepackage{IEEEtrantools}
\usepackage{pdflscape}
\usepackage{afterpage}
\usepackage{capt-of}
\usepackage{graphicx}
\usepackage{enumitem}  

\numberwithin{equation}{section}

\newcommand{\be}{\begin{equation}}
\newcommand{\ee}{\end{equation}}
\newcommand{\non}{\nonumber}

\newcommand{\M}{\mathbb{M}}

\newcommand{\R}{\mathbb{R}}
\newcommand{\T}{\mathbb{T}}

\newcommand{\tr}{\mathop{\rm tr}\nolimits}

\begin{document}

\begin{titlepage}
\strut\hfill UMTG--305
\vspace{.5in}
\begin{center}

\LARGE Wronskian-type formula for inhomogeneous TQ-equations\\ 
\vspace{1in}
\large Rafael I. Nepomechie\footnote{nepomechie@miami.edu}\\
Physics Department, P.O. Box 248046\\
University of Miami, Coral Gables, FL 33124\\[0.8in]
\end{center}

\vspace{.5in}

\begin{abstract}
	The transfer-matrix eigenvalues of the isotropic open Heisenberg
	quantum spin-1/2 chain with non-diagonal boundary magnetic fields
	are known to satisfy a TQ-equation with an inhomogeneous term.  We
	derive here a discrete Wronskian-type formula relating a solution
	of this inhomogeneous TQ-equation to the corresponding solution of
	a dual inhomogeneous TQ-equation.
\end{abstract}

\end{titlepage}

\setcounter{footnote}{0}

\section{Introduction and summary}\label{sec:intro}

The celebrated Baxter TQ-equation for the closed periodic XXX spin 
chain of length $N$\footnote{We use throughout the convenient 
notation $f^{\pm}(u)=f(u\pm\frac{i}{2})$ and $f^{\pm\pm}(u)=f(u\pm i)$.}
\be
T(u)\, Q(u)=(u^{+})^{N} Q^{--}(u) + (u^{-})^{N} Q^{++}(u)
\label{TQclosed}
\ee
can be regarded, for a given transfer-matrix eigenvalue $T(u)$,
as a second-order finite-difference equation for $Q(u)$. The 
eigenvalue $T(u)$ is necessarily a polynomial in $u$ (of degree $N$), since the model is integrable.
Eq.  (\ref{TQclosed}) is well-known to have two independent polynomial solutions
\cite{Pronko:1998xa}: a polynomial $Q(u)$ of degree $M \le N/2$
\be
Q(u) = \prod_{k=1}^{M}(u-u_{k})\,,
\ee
whose zeros $\{ u_{k} \}$ are solutions of the Bethe 
equations that follow directly from (\ref{TQclosed})
\be 
\left(\frac{u_{j}+\tfrac{i}{2}}{u_{j}-\tfrac{i}{2}}\right)^{N}=
\prod_{k=1; k\ne j}^{M}\frac{u_{j}-u_{k}+i}{u_{j}-u_{k}-i} \,, \qquad j = 1, 
\ldots, M \,;
\label{BAE}
\ee
and a second polynomial $P(u)$ of degree $N-M+1 > N/2$
corresponding to Bethe roots ``on the other side of the equator.''
These two solutions are related by a discrete Wronskian (or
Casoratian) formula
\be
P^{+}(u)\, Q^{-}(u) - P^{-}(u)\, Q^{+}(u) \propto u^{N} \,,
\label{Wclosed}
\ee
where $\propto$ denotes equality up to a multiplicative constant.
The existence of a second polynomial solution of the TQ-equation is equivalent to the 
admissibility of the Bethe roots \cite{Mukhin:2009, Tarasov:2018}.
Using the Wronskian formula,
the Q-system for this model \cite{Marboe:2016yyn} (which provides an 
efficient way of computing the admissible Bethe roots) can be succinctly 
reformulated in terms of Q and P \cite{Granet:2019knz, Bajnok:2019zub}.

A generalization of the Wronskian formula (\ref{Wclosed}) 
for the open XXX spin chain with diagonal boundary fields was recently  
obtained \cite{Nepomechie:2019gqt}
\be
g(u)\, P^{+}(u)\, Q^{-}(u) - f(u)\, P^{-}(u)\, Q^{+}(u) \propto u^{2N+1} \,,
\label{Wdiag}
\ee
where $Q(u)$ and $P(u)$ are polynomial solutions of a TQ and a dual-TQ 
equation, respectively. Moreover, the functions $f(u)$ and $g(u)$ are given by
\be
f(u) = (u- i \alpha)(u + i \beta) \,, \qquad g(u) = f(-u) = (u+ i 
\alpha)(u - i \beta) \,, \quad \text{[diagonal case]}
\label{fgdiag}
\ee
where $\alpha$ and $\beta$ are boundary parameters. This result was 
used in \cite{Nepomechie:2019gqt} to formulate a Q-system for the 
model.

The main result of this note is 
a further generalization of the Wronskian formula for the case of
\emph{non-diagonal} boundary fields, namely,
\be
g(u)\, P^{+}(u)\, Q^{-}(u) - f(u)\, P^{-}(u)\, Q^{+}(u) = \mu(u)\, u^{2N+1} \,,
\label{Wnondiag}
\ee
where again $Q(u)$ and $P(u)$ are polynomial solutions of a TQ
(\ref{TQ}) and a dual-TQ (\ref{dualTQ}) equation, respectively; 
$f(u)$ and $g(u)$ are now given by (\ref{fg}); and -- most 
importantly -- $\mu(u)$ is a polynomial that satisfies the following 
remarkably simple relation \footnote{After this work was completed, 
we became aware of a similar result for the closed XXX spin chain with a 
non-diagonal twist, see Theorem 4.10 in \cite{Belliard:2018pvg}.}
\be
\mu^{+}(u) - \mu^{-}(u) = \gamma\, u\, \left( Q(u) - P(u) \right)\,.
\label{mudef}
\ee
In other words, $\mu(u)$ is the \emph{discrete integral} of $\gamma\,
u\, \left( Q(u) - P(u) \right)$.  For the diagonal case, $\gamma=0$;
it then follows from (\ref{mudef}) that $\mu(u)$ is constant, and
therefore (\ref{Wnondiag}) reduces to (\ref{Wdiag}).  The appearance
of the nontrivial factor $\mu(u)$ in the Wronskian-type formula
(\ref{Wnondiag})-(\ref{mudef}) is due to the presence of an
inhomogeneous term in the model's TQ-equation (\ref{TQ}) \cite{Cao:2013qxa,
Nepomechie:2013ila, Wang2015}.  We expect that this Wronskian-type
formula will be useful for formulating a Q-system for this model, 
which however remains a challenge.

In Section \ref{sec:TQ}, we first briefly review the construction of the model
and its TQ-equation, and we then obtain a dual TQ-equation.  We derive
the Wronskian-type formula (\ref{Wnondiag})-(\ref{mudef}) in Section
\ref{sec:Wronskian}.

\section{The model and its TQ-equations}\label{sec:TQ}

We consider the isotropic (XXX) open Heisenberg quantum spin-1/2
chain of length $N$ with boundary magnetic fields, whose Hamiltonian
is given by
\be
H = \sum_{k=1}^{N-1} \vec \sigma_{k} \cdot  \vec \sigma_{k+1}
- \frac{\xi}{\beta} \sigma^{x}_{1} - \frac{1}{\beta} \sigma^{z}_{1} 
+ \frac{1}{\alpha} \sigma^{z}_{N} \,, 
\label{Ham}
\ee
where $\vec \sigma = (\sigma^{x}\,, \sigma^{y}\,, \sigma^{z})$ are 
the usual Pauli matrices, and
$\alpha$, $\beta$ and $\xi$ are arbitrary real parameters. 
For the non-diagonal case $\xi \ne 0$, this model is not $U(1)$-invariant.

In order to construct the corresponding transfer matrix, we use the R-matrix
(solution of the Yang-Baxter equation) given by the $4 \times 4$ matrix
\be
\mathbb{R}(u)=(u-\tfrac{i}{2})\mathbb{I}+i\mathbb{P}\,,
\label{Rmat}
\ee
where $\mathbb{P}$ is the permutation matrix and $\mathbb{I}$ is the 
identity matrix;
and the K-matrices (solutions of boundary Yang-Baxter 
equations) given by the $2 \times 2$ matrices \cite{Ghoshal:1993tm, deVega:1993xi}
\begin{align}
\mathbb{K}^{R}(u) &= \begin{pmatrix}
i(\alpha-\tfrac{1}{2}) + u & 0 \\  
0 & i(\alpha+\tfrac{1}{2}) - u
\end{pmatrix}\,, \non \\
\mathbb{K}^{L}(u) &= \begin{pmatrix}
i(\beta-\tfrac{1}{2}) - u\ & -\xi(u+\tfrac{i}{2}) \\ 
-\xi(u+\tfrac{i}{2}) & i(\beta+\tfrac{1}{2}) + u 
\end{pmatrix} \,,
\label{Kmat}
\end{align}
which depend on the boundary parameters $\alpha$, $\beta$ and $\xi$.

The transfer matrix $\T(u)= \T(u; \alpha, \beta, \xi)$ is given by \cite{Sklyanin:1988yz}
\be
\T(u) = \tr_{0} \mathbb{K}^{L}_{0}(u)\, \M_{0}(u)\, 
\mathbb{K}^{R}_{0}(u)\, \widehat{\M}_{0}(u) \,,
\label{transf}
\ee
where $\mathbb{M}$ and $\widehat{\M}$ are monodromy matrices given by
\begin{align}
\mathbb{M}_{0}(u) &=\mathbb{R}_{01}(u)\, 
\mathbb{R}_{02}(u)\dots\mathbb{R}_{0N}(u) \,, \non \\
\widehat{\M}_{0}(u) &= \R_{0 N}(u) \cdots \R_{0 2}(u)\, \R_{0 1}(u) \,.
\end{align}
The transfer matrix is engineered to have the fundamental commutativity property
\be
\left[ \T(u) \,, \T(v) \right] = 0 \,,
\label{commutativity}
\ee
and satisfies $ \T(-u) =  \T(u)$.
The Hamiltonian (\ref{Ham}) is proportional to 
$\frac{d\T(u)}{du}\Big\vert_{u=i/2}$, up to an additive constant. 

The eigenvalues $T(u)$ of the transfer matrix $\T(u)$ are polynomials 
in $u$ (as a consequence of (\ref{commutativity})), and satisfy the 
TQ-equation \cite{Cao:2013qxa, Nepomechie:2013ila, Wang2015}
\be
-u\, T(u)\, Q(u) = g^{-}(u)\,(u^{+})^{2N+1} Q^{--}(u)
+f^{+}(u)\, (u^{-})^{2N+1} Q^{++}(u) 
- \gamma\, u \left( u^{-} u^{+}\right)^{2N+1} \,,
\label{TQ}
\ee
where $Q(u)$ is an even polynomial of degree $2N$
\be
Q(u) = \prod_{k=1}^{N}(u-u_{k})(u+u_{k})\,,
\ee
the functions $f(u)$ and $g(u)$ are given by
\be
f(u) = (u- i \alpha)\left(u\sqrt{1+\xi^{2}}  + i \beta \right) \,, \qquad 
g(u) = f(-u) = (u+ i \alpha)\left(u\sqrt{1+\xi^{2}} - i \beta \right) \,,
\label{fg}
\ee
and we define $\gamma$ by
\be
\gamma = -2\left(1-\sqrt{1+\xi^{2}}\right) \,.
\label{gamma}
\ee
Note the presence of an inhomogeneous term (proportional to $\gamma$) 
in the TQ-equation (\ref{TQ}).
For the diagonal case $\xi = 0$, we see from (\ref{gamma}) that 
$\gamma=0$, hence the inhomogeneous term disappears; moreover, the functions
$f(u)$ and $g(u)$ (\ref{fg}) reduce to (\ref{fgdiag}).

The transfer matrix transforms under charge conjugation by reflection (negation) of all the 
boundary parameters
\be
{\cal C}\, \T(u; \alpha, \beta, \xi) \, {\cal C} = \T(u; -\alpha, 
-\beta, -\xi) \,, \qquad {\cal C} = (\sigma^{x})^{\otimes N}
\,.
\ee
We therefore obtain a dual TQ-equation from (\ref{TQ})
as in \cite{Nepomechie:2019gqt} by making the
replacement $Q(u) \mapsto P(u)$, and by reflecting the boundary 
parameters $(\alpha \mapsto - \alpha\,, \beta \mapsto - \beta\,, \xi 
\mapsto - \xi)$, which implies that $f(u)$ and $g(u)$ become interchanged
\be
-u\, T(u)\, P(u) = f^{-}(u)\,(u^{+})^{2N+1} P^{--}(u)
+ g^{+}(u)\, (u^{-})^{2N+1} P^{++}(u) 
- \gamma\, u \left( u^{-} u^{+}\right)^{2N+1} \,,
\label{dualTQ}
\ee
where $P(u)$ is also an even polynomial of degree $2N$
\be
P(u) = \prod_{k=1}^{N}(u-\tilde{u}_{k})(u+\tilde{u}_{k})\,,
\ee
whose zeros can be regarded as dual Bethe roots.
We emphasize that the same eigenvalue $T(u)$ appears in both 
(\ref{TQ}) and (\ref{dualTQ}).

\section{The Wronskian-type formula}\label{sec:Wronskian}

We now look for a relation between $Q(u)$ (a solution of the
TQ-equation (\ref{TQ}) for some transfer-matrix eigenvalue $T(u)$) and
the corresponding $P(u)$ (a solution of the dual TQ-equation
(\ref{dualTQ}) for the same transfer-matrix eigenvalue $T(u)$).  To
this end, we make the ansatz
\be
g(u)\, P^{+}(u)\, Q^{-}(u) - f(u)\, P^{-}(u)\, Q^{+}(u) = \mu(u)\, u^{2N+1} \,,
\label{ansatz}
\ee
where $f(u)$ and $g(u)$ are given by (\ref{fg}), and the function $\mu(u)$ is still to be determined.
This ansatz is motivated by the result (\ref{Wdiag}) for the diagonal 
case. 

We next define $R(u)$, along the lines of \cite{Pronko:1998xa}, by
\be
R = \frac{u^{2N+1}}{Q^{+}\, Q^{-}} 
= \frac{1}{\mu}\left(g \frac{P^{+}}{Q^{+}} - f \frac{P^{-}}{Q^{-}} 
\right) \,,
\label{R}
\ee
where the second equality follows from the ansatz (\ref{ansatz}). 
Dividing both sides of the TQ-equation (\ref{TQ}) by $Q\, Q^{++}\, 
Q^{--}$, we obtain
\be
-\frac{u\, T}{Q^{++}\, Q^{--}} = f^{+} R^{-} + g^{-} R^{+} - \gamma\, 
u\, Q\, R^{+} R^{-} \,.
\label{TQR}
\ee
Substituting in (\ref{TQR}) for $R$ using the second equality in (\ref{R}), and then
multiplying both sides by $\mu^{+} \mu^{-} Q^{++} Q^{--}$, we obtain
\begin{align}
-u\, T\, \mu^{+} \mu^{-} 
&=f^{+} g^{-} \frac{P\, Q^{++} Q^{--}}{Q}(\mu^{+} - \mu^{-} + 
\gamma\, u\, P) - f^{+} f^{-} P^{--} Q^{++} (\mu^{+} + \gamma\, u\, 
P) \non\\
&\quad + g^{+} g^{-} P^{++} Q^{--} (\mu^{-} - \gamma\, u\, P) + f^{-} g^{+} 
P^{--} P^{++} Q\, (\gamma\, u) \,.
\label{resultR}
\end{align} 
Similarly, we define $S(u)$ by
\be
S = \frac{u^{2N+1}}{P^{+}\, P^{-}} 
= \frac{1}{\mu}\left(g \frac{Q^{-}}{P^{-}} - f \frac{Q^{+}}{P^{+}} 
\right) \,,
\label{S}
\ee
and we divide both sides of the dual TQ-equation (\ref{dualTQ}) by 
$P\, P^{++}\, P^{--}$, thereby obtaining
\be
-\frac{u\, T}{P^{++}\, P^{--}} = f^{-} S^{+} + g^{+} S^{-} - \gamma\, 
u\, P\, S^{+} S^{-} \,.
\label{TQS}
\ee
Substituting in (\ref{TQS}) for $S$ using the second equality in (\ref{S}), and then
multiplying both sides by $\mu^{+} \mu^{-} P^{++} P^{--}$, we obtain
\begin{align}
-u\, T\, \mu^{+} \mu^{-} 
&=f^{+} g^{-} P\, Q^{++} Q^{--} (\gamma\, u) - f^{+} f^{-} P^{--} 
Q^{++} (\mu^{-} + \gamma\, u\, Q) \non\\
&\quad + g^{+} g^{-} P^{++} Q^{--} (\mu^{+} - \gamma\, u\, Q) 
- f^{-} g^{+} \frac{P^{--} P^{++} Q}{P}\, (\mu^{+} - \mu^{-} 
-\gamma\, u\, Q) \,.
\label{resultS}
\end{align} 
Equating the right-hand-sides of (\ref{resultR}) and (\ref{resultS}), 
we arrive at the constraint
\be
\left[\mu^{+} - \mu^{-} + \gamma\, u\, (P - Q) \right] 
\frac{1}{Q P}\left(g P^{+}Q^{-} - f P^{-} Q^{+}\right)^{+} 
\left(g P^{+}Q^{-} - f P^{-} Q^{+}\right)^{-} = 0 \,.
\ee
This constraint can evidently be satisfied by setting
\be
\mu^{+} - \mu^{-} = \gamma\, u\, (Q - P) \,,
\label{mudef2}
\ee
as claimed in (\ref{mudef}). For given polynomials $Q(u)$ and $P(u)$,
(\ref{mudef2}) can be solved for a polynomial function $\mu(u)$, up 
to an arbitrary additive constant.

The result (\ref{mudef2}) can in fact be obtained in a more 
straightforward way:\footnote{The author thanks the Referee for this 
nice observation.} 
define $\mu(u)$ by
\be
\mu(u) = \frac{1}{u^{2N+1}}\left(g(u)\, P^{+}(u)\, Q^{-}(u) - f(u)\, P^{-}(u)\, 
Q^{+}(u)\right) \,, 
\label{mudef3}
\ee
which is equivalent to (\ref{Wnondiag}). Multiply the TQ-equation (\ref{TQ}) by 
$\frac{P(u)}{(u^{-} u^{+})^{2N+1}}$, multiply the dual TQ-equation 
(\ref{dualTQ}) by $\frac{Q(u)}{(u^{-} u^{+})^{2N+1}}$, and subtract 
the second equation from the first. The result, when expressesed in 
terms of $\mu$ (\ref{mudef3}), is exactly (\ref{mudef2}).

\section*{Acknowledgments}

The author thanks the organizers of the CQIS-2019 workshop in St. Petersburg for 
their kind invitation. He was supported in part by a Cooper fellowship.

\noindent
Conflict of Interest: The author declares that he has no conflict of interest.


\providecommand{\href}[2]{#2}\begingroup\raggedright\endgroup

\end{document}